\documentclass[aip,jcp,a4paper,twocolumn]{revtex4}

\usepackage[]{amsmath}
\usepackage{amssymb}
\usepackage[dvips]{graphicx}
\usepackage{color}
\usepackage{tabularx}
\usepackage{mathtools}
\usepackage{algpseudocode}
\usepackage{enumitem}
\usepackage{multirow}
\usepackage{epstopdf}  
\usepackage{bm}
\usepackage{tikz}
\usetikzlibrary{shapes,snakes}
\usetikzlibrary{arrows}


\newcommand{\recheck}[1]{{#1}}

\newcommand{\vect}[1]{\textbf{\textit{#1}}}

\newcommand{\out}{{\mathrm{out}}}
\newcommand{\bias}{{\mathrm{bias}}}

\newcommand{\me}{\mathcal {E}}

\newcommand{\ml}{\mathcal {L}}
\newcommand{\ma}{\mathcal {A}}

\newcommand{\mf}{\boldsymbol{\mathcal {F}}}

\newcommand{\newr}{\mathrm {accept}}

\newcommand*\chem[1]{\ensuremath{\mathrm{#1}}}

\begin{document}

\title{
 Reinforced dynamics for enhanced sampling in large atomic and molecular systems
  }
\author{Linfeng Zhang}
\affiliation{Program in Applied and Computational Mathematics, 
Princeton University, Princeton, NJ 08544, USA}
\author{Han Wang}
\email{wang\_han@iapcm.ac.cn}
\affiliation{Institute of Applied Physics and Computational Mathematics,
Fenghao East Road 2, Beijing 100094, P.R.~China}
\affiliation{CAEP Software Center for High Performance Numerical
Simulation, Huayuan Road 6, Beijing 100088, P.R.~China}
\author{Weinan E}
\email{weinan@math.princeton.edu}
\affiliation{Department of Mathematics and Program 
in Applied and Computational Mathematics, 
Princeton University, Princeton, NJ 08544, USA}
\affiliation{Beijing Institute of Big Data Research, 
Beijing, 100871, P.R.~China}

\begin{abstract}
    A new approach for efficiently exploring the configuration space and computing the free energy of large atomic and molecular systems 
    is proposed, motivated by an analogy with reinforcement learning. There are two major components in this new approach.
    Like metadynamics, it allows for an efficient exploration of the configuration space by adding an adaptively computed biasing
  potential to the original dynamics.  Like deep reinforcement learning, this biasing potential is trained on the fly using deep
  neural networks, with data collected judiciously from the exploration and an uncertainty indicator from the neural network model playing the role of the reward function. 
  Parameterization using neural networks makes it feasible to handle cases with a large set of collective variables.
  This has the potential advantage that selecting precisely the right set of collective variables has now become less critical for
  capturing the  structural transformations of the system.
  The method is illustrated by studying the full-atom, explicit solvent models 
  of alanine dipeptide and tripeptide, as well as the system of a polyalanine-10 molecule with 20 collective variables.
\end{abstract}

\maketitle

\setlength{\parskip}{.5em}

\section{Introduction}

Exploring the configuration space of large atomic and molecular systems is a problem of fundamental importance for
many applications, including protein folding, materials design, and understanding chemical reactions, etc.
There are several difficulties associated with these applications. 
The first is that the dimensionality of the configuration space is typically very high.
The second is that there are often high energy barriers associated with the exploration.
Both difficulties can be reduced by the introduction of collective variables (CVs) and the mapping of the problem
to the CV space. The problem then becomes finding the free energy surface (FES) associated with the set of CVs,
a problem that has attracted a great deal of interest in the last few decades
\cite{kumar1992weighted,kumar1995multidimensional,voter1997hyperdynamics,sugita1999replica,vandevondele2000efficient,earl2005parallel,
christ2008multiple,gao2008self,laio2002escaping,barducci2008well,maragliano2006temperature,maragliano2008single,abrams2008efficient,
yu2011temperature}.
One of the most effective techniques is metadynamics~\cite{laio2002escaping},
which computes a biasing potential by depositing Gaussian bases along the trajectory in the CV space.
It is shown that the biasing potential converges to the inverted free energy at the end of the calculation~\cite{barducci2008well}.
Also closely related to our work are the recent papers that propose to use machine learning methods to
help parameterizing FES \cite{stecher2014free,mones2016exploration,galvelis2017neural,schneider2017stochastic}.
In particular, the deep neural network (DNN) model has shown promise in effectively representing the FES defined on high dimensional CV space~\cite{galvelis2017neural,schneider2017stochastic}.

In this work, we take metadynamics and machine learning methods one step further
by making an analogy between reinforcement learning \cite{sutton1998reinforcement}
and the task of configuration space exploration and FES calculation.
Classical reinforcement learning scheme involves a state space, an action space, and a reward function. The objective is to find the
best policy function, which is a mapping from the state space to the action space, that optimizes the cumulative reward function.
Our problem can be thought of as being a multi-scale reinforcement learning problem.
We have a micro-state space, the configuration space of the detailed atomic system, and a macro-state space,
the space of the CVs. The action space will be represented by the biasing potential in the biased molecular dynamics 
on the micro-state space. The optimal policy function is the inverted FES, defined on the macro-state space. 
The FES is parameterized by a carefully designed DNN model.
Among other things, this allows us to handle cases with a large set of CVs.
In the absence of an explicit reward function, we introduce an uncertainty indicator that can be used to quantify the
accuracy of the FES representation.
It is defined as the standard deviation of the predictions from an ensemble of DNN models,
which are trained using the same dataset but different initialization of the model parameters.
The bias is only adopted in regions where the uncertainty indicator is low, i.e. regions that are sufficiently explored, 
and thus the exploration in the insufficiently explored region is encouraged.
We call this scheme the ``reinforced dynamics'', to signal its analogy with reinforcement learning.

Roughly speaking, reinforced dynamics works as follows:
The biasing potential, or the action, is initialized at 0 and is expected to converge to the inverted FES as the
dynamics proceeds. Each step of the macro-iteration involves the following components.
First, a biased MD is performed, in which the system is biased only in the regions where the uncertainty indicator is low.
The biased simulation is likely to visit the CV regions never visited before or where the FES representation quality is poor.
Next, a certain number of the newly visited CV values  in  regions where the uncertainty indicator is high are added to the training dataset.
A restrained MD is performed to obtain the mean force, or the negative gradient of the FES,
at each of the newly added CV values.
Finally, the accumulated CV values and the mean forces are used as labels to train several network models,
which give the current estimate of the biasing potential and the uncertainty indicator.
This process is repeated iteratively until convergence is achieved,
when the newly visited CV values all fall in the regions where the uncertainty indicator is low.

The quality of the free energy surface is determined by the quality of the CVs.
Ideally we would like the FES to capture the structural and dynamic information of the system, such as
the important metastable states and transitions between the metastable states.
For many years, since our ability to accurately approximate the FES has  been limited to systems with a small
number of CVs, we have always faced the dilemma that choosing the right CVs is both critical and
practically impossible.  
We believe that the ability of the reinforced dynamics to handle a large set of CVs will make
the issue of choosing the right CVs much less critical.

In this paper, we give a systematic presentation of the theoretical and practical aspects of reinforced dynamics.
We first focus on methodology and introduce the theory and flowchart of the reinforced dynamics scheme. 
Then we use the classical example of alanine dipeptide and tripeptide with two and four CVs, respectively, as illustrations due to their intuitive appeal.  
{The solvent effect is explicitly considered in both examples.}
The FESs constructed by the reinforced dynamics are compared with those constructed by long brute-force simulations (5.1~$\mu$s for alanine dipeptide and 47.7~$\mu$s for tripeptide) to demonstrate the accuracy and efficiency of the method. 
Finally, an application to the structural optimization of the polyalanine-10 system with 20 CVs is presented to demonstrate  the practical promise of reinforced dynamics.

\section{Theory}
\subsection{Free energy and mean forces}
We assume that the system we are studying has $N$ atoms, with their positions denoted by
$\bm r = (\bm r_1, \dots, \bm r_N)$.
The potential energy of the system is denoted by $U(\bm r)$.
Without loss of generality, we consider the system in a canonical ensemble.
Given  $M$ predefined CVs, denoted by $\bm s(\bm r) = (s_1(\bm r), \dots, s_M(\bm r))$,
the free energy defined on the CV space is
\begin{align}\label{eqn:fe}
  A(\bm s) = -\frac{1}\beta
  \ln p(\bm s), \quad
  p(\bm s)=
  \frac 1{Z} \int e^{-\beta U (\bm r)} \delta (\bm s(\bm r) - \bm s)\, d\bm r,
\end{align}
with $Z = \int e^{-\beta U (\bm r)}\,d\bm r$ being the normalization factor.
The brute-force way of computing the free energy~\eqref{eqn:fe} is to sample the CV
space exhaustively
and to approximate the probability distribution $p (\bm s)$ by making a histogram of the CVs.
This approach may easily become prohibitively expensive.
In such a case, an alternative  way of constructing the FES is to fit the mean forces acting on the CVs, i.e.,
\begin{align}\label{eqn:mf}
  \bm F (\bm s) = -\nabla_{\bm s} A(\bm s).
\end{align}
Several ways of computing  $\bm F (\bm s)$ have been proposed \cite{ciccotti2005blue,maragliano2006temperature,abrams2008efficient}.
We will adopt the approach of restrained dynamics proposed  in
~\cite{maragliano2006temperature}.
In this formulation, a new term is added to the potential of the system to represent the effect of the 
spring forces between the configuration variables and the CVs.  It can be shown that  the
mean force is given by
$F_{\alpha} (\bm s) = \lim_{k_\alpha \rightarrow\infty} F_{\alpha}^k (\bm s)$ for $\alpha=1,2,...,M$,
where  the $\alpha$-th component of $F^k$ is defined to be
\begin{align}\label{eqn:mf-res}
  F_\alpha^k(\bm s) =
  \frac1{Z_k(\bm s)}
  \int
  k_\alpha (s_\alpha(\bm r) - s_\alpha)\, e^{-\beta U_k (\bm r, \bm s)} 
  \,d\bm r.
\end{align}
Here  $Z_k(\bm s) = \int e^{-\beta U_k (\bm r,\bm s)}\,d\bm r$ is the normalization factor, 
$\{k_\alpha \,\vert\, \alpha = 1,\dots,M\}$ are the spring constants for the harmonic restraining potentials,
and $U_k (\bm r, \bm s)$ is defined by
\begin{align}
  U_k (\bm r, \bm s) = U(\bm r) +
  \sum_{\alpha=1}^M
  \frac12 k_\alpha ( s_\alpha (\bm r) - s_\alpha)^2.
\end{align}

In practice, the spring constants are chosen to be large enough to guarantee the convergence to the mean forces.
The time duration for the restrained dynamics
should be longer than the largest relaxation timescale of the fast modes of the system,
in order for  the ensemble average in Eq.~\eqref{eqn:mf-res} to be approximated adequately  by the time average.
In the rest of the paper, we do not explicitly distinguish $\bm F$ and $\bm F^k$.

\subsection{Free energy representation}

\begin{figure}
  \centering
  \includegraphics[width=0.45\textwidth]{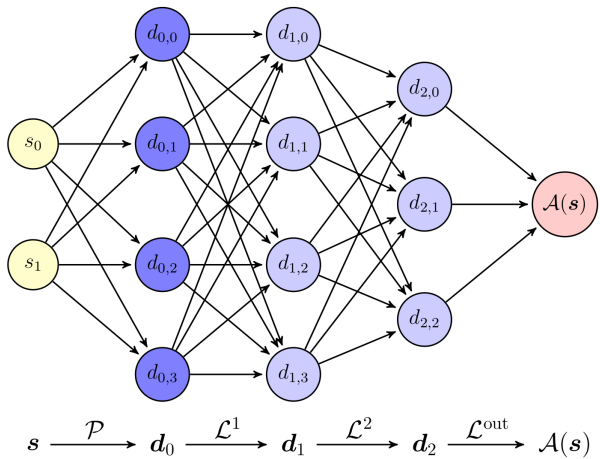}
  \caption{A schematic plot of the DNN representation of the free energy $A(\bm s)$.
    As an example,
    the dimension of the CV space in the figure is $M=2$.
    The preprocessing operator $\mathcal P$ maps the CV values to an input layer that has $M_0 = 4$ nodes. 
    The DNN has $2$ hidden layer, namely $\bm d_1$ and $\bm d_2$, the size of which are $M_1 = 4$ and $M_2 = 3$, respectively.
    The last hidden layer $\bm d_2$ is mapped to the free energy $A(\bm s)$ by the output operator $\ml^\out$.
  }
  \label{fig:dnn}
\end{figure}

The free energy $A(\bm s)$ will be represented by a deep neural network (DNN) model, in which
the input CVs are first preprocessed, then passed through multiple fully connected hidden layers,
and, in the end, mapped to the free energy.
The structure of the DNN model is schematically illustrated in Fig.~\ref{fig:dnn}.
Mathematically, a DNN representation with $N_h$ hidden layers is given by 
\begin{align}\label{eqn:dnn}
  \ma (\bm s) = \ml^\out\circ\ml^{N_h}\circ\cdots\circ\ml^{1}\circ\mathcal P (\bm s)
\end{align}
where ``$\circ$'' denotes function composition.
The differentiable operator $\mathcal P$ represents the system-dependent
preprocessing procedure for the CVs, which will be illustrated by the examples in Sec.~\ref{sec:example}. 
For the $p$-th hidden layer, which has $M_p$ neurons $\bm d_p \in\mathbb R^{M_p}$, 
$\ml_p$ is the operation that maps $\bm d_{p-1}$ to $\bm d_p$, using:
\begin{align}\label{eqn:weights}
  \bm d_p = \ml_p (\bm d_{p-1}) = \varphi (\bm W_p \bm d_{p-1} + \bm b_p).
\end{align}
Here $\bm W_p\in \mathbb R^{M_p\times M_{p-1}}$ and $\bm b_p \in \mathbb R^{M_p}$ are
coefficients of a linear mapping, often called weights.   $\varphi$ is the so-called activation
function, which is in general nonlinear.
In this project  we use the component-wise hyperbolic tangent function for $\varphi$.
The output layer $\ml^\out$ is defined by
\begin{align}
  \ma(\bm s) = \ml_\out(\bm d_{N_h}) = \bm W_\out \cdot \bm d_{N_h-1} + b_\out
\end{align}
where $\bm W_\out\in \mathbb R_{ M_{p-1}}$ and $b_\out \in \mathbb R$ are the weights 
of the linear mapping.
Finally, $\bm W = \{\bm W_1, \dots, \bm W_{N_h}, \bm W_\out\}$ and
$\bm b = \{\bm b_1, \dots, \bm b_{N_h}, b_\out \}$ constitute all the DNN model parameters to be determined.
We note that the gradient, representing the mean force
\begin{align}
  \mf (\bm s) = -\nabla_{\bm s}\ma (\bm s)
\end{align}
is well defined since
each layer of the construction~\eqref{eqn:dnn} is differentiable,
and hence the DNN representation of the free energy $\ma(\bm s)$ is also differentiable.

It should be noted that the design of the DNN model can be adapted to different kinds of problems.
We use the fully-connected DNN model here  for simplicity of discussion.
For example, for some condensed systems, an alternative network model
resembling the one used in the Deep Potential method should be preferred \cite{han2017deep,zhang2017deep}. 
We leave this to future work.

\subsection{Training and uncertainty indicator}

The DNN representation of the free energy is obtained by solving the following
minimization problem 
\begin{align}\label{eqn:min}
  \min_{\{\bm W,\bm b\}} L_D(\{\bm W,\bm b\}) .
\end{align}
The loss function $L_D$ is defined by
\begin{align}\label{eqn:loss-approx}
  L_D(\{\bm W,\bm b\})=
  \frac 1{\vert D\vert}
  \sum_{\bm s \in D}
  \Vert
  \mf(\bm s) - \vect F(\bm s)
  \Vert^2,
\end{align}
where $D$ denotes the set of training data and
$\vert D\vert$ denotes the size of the dataset $D$.
Here $\mf(\bm s)$ comes from the DNN model, and $F(\bm s) $ is the collected mean force for the data $s$.
Precise ways of collecting the data  will be discussed later.
It should be noted that at the beginning of the training process, we have no  data.
 Data is collected as the training process proceeds.

To guarantee accuracy for this model, we require that 
the CV values in $D$ is an adequate  sample  of the CV space.
This is made difficult due to the barriers on the energy landscape.
The MD will tend to be stuck at metastable states without being able to escape.
To help overcome this problem, we introduce a biased dynamics. Details of that will be discussed in the next subsection.

A key notion for reinforced dynamics is the uncertainty indicator.  This quantity is
important in the data collection step as well as in the biased dynamics step. 
Our intuition  is that
  the DNN model should produce a reasonably accurate prediction of the free energy in  
regions that are adequately covered by $D$,
but is much less so in  regions that are covered poorly by $D$ (or have not been visited by the MD).
To quantify this, we introduce a small ensemble of DNN models, where the only difference between
these models is the random weights used to initialize them.  We can then define 
the
uncertainty indicator as $\me(\bm s)$, the standard deviation of the force predictions, viz.
\begin{align}\label{eqn:model-std}
  \me^2 (\bm s) =
  \big\langle
  \Vert
  \mf (\bm s) - \bar\mf(\bm s)
  \Vert^2
  \big\rangle,
  \quad
  \bar\mf(\bm s) =
  \big\langle
  \mf (\bm s)  
  \big\rangle,
\end{align}
where the ensemble average $\langle\cdots\rangle$ is taken over
this ensemble of models.
One expects that this ensemble of models
 give rise to  predictions of the mean forces $\bm F$  that are close to each other in regions well covered by $D$.
In the regions that are covered poorly by $D$, the predictions will scatter much more.
This is confirmed by our numerical results. 

Finally, it is worth noting that
the minimization problem~\eqref{eqn:min} is solved by the stochastic gradient descent (SGD) method
combined with the back-propagation algorithm~\cite{lecun2012efficient}.
This has become the \emph{de facto} standard algorithm for training DNN models.
\recheck{
In all the test examples, we first adopt a random initialization 
procedure for the weights, 
where each component in $\bm W_p $ in Eq.~\eqref{eqn:weights} is initialized from a normal distribution with mean 0 and standard deviation ${1}/{\sqrt{|\bm d_{p-1}|+|\bm d_p|}}$,
and each component in $\bm b_p $ is initialized from a 
normal distribution with mean 0 and standard deviation 1.}
Then at each training step, the weights are updated based on the evaluation of the loss function on a small \emph{batch}, 
or subset $B$ of the training data $D$, i.e.,
\begin{align}
  L = \frac{1}{\vert B\vert}
  \sum_{\bm s \in B}
  \big\Vert
  \mf(\bm s) - \vect F(\bm s)
  \big\Vert^2.
\end{align}

\subsection{Adaptive biasing}

A way of encouraging the MD to overcome the barriers in the energy landscape and escape metastable regions is to add
a bias to the potential. The force on the $i$-th atom then becomes:
\begin{align}
 \tilde {\bm {f}}_i(\bm r) = -\nabla_{\bm r_i} U(\bm r) -\nabla_{\bm r_i} U_\bias(\bm s(\bm r)).
\end{align}
Since the FES is the best approximation of the potential energy in the space of CVs, it is natural to use
  the current approximation of the FES, with a negative sign added, as the  biasing potential, as is done
in metadynamics \cite{laio2002escaping,barducci2008well}.
We will adopt the same strategy but we propose to switch on the biasing potential 
only in regions where we have low uncertainty on the DNN representation of the FES:
\begin{align}\label{eqn:ab}
  \tilde {\bm {f}}_i(\bm r) = -\nabla_{\bm r_i} U(\bm r) + \sigma(\me(\bm s(\bm r))) \,\nabla_{\bm r_i}\ma (\bm s(\bm r)),
\end{align}
where the biasing potential $\ma (\bm s(\bm r))$ is the mean of the predefined ensemble of DNN models, and
$\sigma(\cdot)$  is a smooth switching function defined by
\begin{align}
  \sigma(e) =
  \left \{
  \begin{aligned}
    &1, & \quad  & e < e_{0}, \\
    &\frac12 + \frac12 \cos \Big(\pi \, \frac{e - e_{0}}{e_{1} - e_{0}} \Big), & \quad &e_{0}\leq e < e_{1},  \\
    &0,  & \quad  &e \geq e_{1}.
  \end{aligned}
  \right.
\end{align}
Here $e_0$ and $e_1$ are two uncertainty levels for the accuracy of the DNN model.
In regions where the uncertainty indicator $\me(\bm s)$ is smaller than the level $e_0$,
the accuracy of the DNN representation of  $\ma(\bm s)$ is adequate, 
and hence the system will be biased by $\ma(\bm s)$.
In the regions where $\me(\bm s)$ is larger than level $e_1$,
the accuracy of the DNN representation is inadequate,
and the system will follow the original dynamics governed by the potential energy $U(\bm r)$. 
In between $e_0$ and $e_1$, the  DNN model is partially used to bias the system via
a rescaled force term $- \sigma(\me(\bm s(\bm r))) \,\nabla_{\bm r_i}\ma (\bm s(\bm r))$.

\subsection{Data collection}

After the biased MD, a number of the newly visited CV values that are in the regions with high uncertainty
are added to the training dataset $D$.
The regions with high uncertainty are defined to be the CV values that give rise to large uncertainty indicator,
viz.,~$ \me (\bm s) > e_\newr $.
A reasonable choice of the threshold is $e_\newr = e_0$.
For each value of the CV in $D$, we use the restrained dynamics to calculate 
the mean forces $\bm F$ via 
Eq.~\eqref{eqn:mf-res}.
These values, together with those computed in previous iterations, are used as the labels for training the next updated model.

\subsection{The reinforced dynamics scheme}

\begin{figure}
  \centering
  \includegraphics[width=0.48\textwidth]{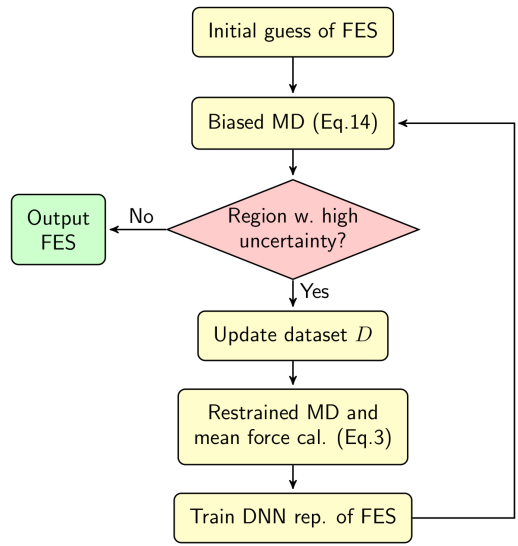}
  \caption{The flowchart of the reinforced dynamics scheme.}
  \label{fig:flowchart}
\end{figure}

 Fig.~\ref{fig:flowchart} is a flowchart of the reinforced dynamics scheme.
Given an initial guess of the FES represented by the DNN, a biased MD, i.e.~Eq.~\eqref{eqn:ab}, 
is performed to sample the CV space from an arbitrarily chosen starting point.
If no \emph{a priori} information on the FES is available, then a standard MD is carried out.
The visited CV values  are recorded at a certain time interval and tested by the 
uncertainty indicator to see whether they
belongs to a region with high uncertainty in the CV space.
If all the newly sampled CV values from the biased MD trajectory belong to the region with low uncertainty,
it can be
(1) the biased MD is not long enough, so  parts of the CV space are not explored, 
(2) the  interval for recording CV values along the biased MD is not small enough, so some visited CV values
 belonging to the region with high uncertainty are missed,
or
(3) the DNN representation for FES is fully converged, then the iteration should be stopped and one can output the
 DNN representation for the FES, namely the mean of the predefined ensemble of models.
Case~(1) can be excluded by systematically increasing the length of the biased simulation.
Case~(2) can be excluded by decreasing the recording interval.

If CV values belonging to the region with high uncertainty are discovered,
they will be added to the training dataset $D$.
The CV values that are already in the training dataset should be retained
and serve as training data for later iterations.
The mean forces
at the added CV values are computed by the restrained dynamics Eq.~\eqref{eqn:mf-res}.
A new ensemble of DNN models  for the FES are then trained,
using different random initial guesses for $\{\bm W, \bm b\}$.
The standard deviation of the predictions from these models is again
 used to estimate the uncertainty indicator $\me (\bm s)$.
The iteration starts again using the biased MD simulation with the new DNN models.

Finally, it is worth noting that the restrained MD simulations for mean forces, which take over most of the computation time in the reinforced dynamics scheme, are embarrassingly parallelizable. 
The training of the ensemble of DNN models is also easily parallelizable.
Several independent walkers can be set up simultaneously
for a parallelized biased simulation, and this  provides a more efficient exploration of the FES.
These techniques can help accelerating the data collection process and benefit large-scale simulations for complex systems.

\section{Numerical examples: alanine dipeptide and tripeptide}\label{sec:example}

\subsection{Simulation setup}
\label{sec:example-ala-2}

We investigate the FES of the alanine dipeptide (ACE-ALA-NME)
and alanine tripeptide (ACE-ALA-ALA-NME)
modeled by the Amber99SB force field~\cite{hornak2006comparison}. 
The molecules are dissolved in 342 and 341 TIP3P~\cite{jorgensen1983comparison} water molecules, respectively, in a periodic simulation cell.
All the MD simulations are performed using the package GROMACS~5.1.4~\cite{abraham2015gromacs}.
The cut-off radius of the van der Waals interaction is 0.9~nm.
The dispersion correction due to the finite cut-off radius is applied to both energy and pressure calculations.
The Coulomb interaction is treated with smooth particle mesh Ewald method~\cite{essmann1995spm}
with a real space cut-off 0.9~nm and reciprocal space grid spacing 0.12~nm.
The system is integrated with the leap-frog scheme at timestep 2~fs.
The temperature of the system is set to 300~K by velocity-rescale thermostat~\cite{bussi2007canonical} with a relaxation time 0.2~ps.
The solute and solvent are coupled to two independent thermostats to avoid the hot-solvent/cold-solute problem~\cite{lingenheil2008hot}.
Parrinello-Rahman barostat~\cite{parrinello1981polymorphic} (GROMACS implementation)
with a relaxation timescale 1.5~ps and compressibility $4.5\times 10^{-5}\,\textrm{Bar}^{-1}$ is coupled to the system to control the pressure to 1~Bar.
{For both the alanine dipeptide and tripeptide, any covalent bond that connects a hydrogen atom is constrained by the LINCS algorithm~\cite{hess1997lincs}.}
The H-O bond and H-O-H angle of water molecules are constrained by the SETTLE algorithm~\cite{miyamoto2004settle}.

For the alanine dipeptide, two torsion angles $\varphi$ (C, N, \chem{C_\alpha}, C) and
$\psi$ (N, \chem{C_\alpha}, C, N),
are chosen as CVs for this system, i.e.~$\bm s = (\varphi,\psi)$.
While for the alanine tripeptide, the same torsion angles associated with the first and second \chem{C_\alpha}s, denoted by 
$\varphi_0$, $\psi_0$, and $\varphi_1$, $\psi_1$, respectively, 
are used as CVs for the system, i.e.~$\bm s = (\varphi_0, \psi_0, \varphi_1, \psi_1)$.
The GROMACS source code is 
modified and linked to PLUMED~2.4b~\cite{tribello2014plumed} to 
carry out the biased and restrained simulations.
The PLUMED package is modified to compute the DNN biasing force, viz.,~Eq.~\eqref{eqn:ab}.
The DNN models used in both examples contain three hidden layers of size $(M_1, M_2, M_3) = (48, 24, 12)$.
The preprocessing operator for the alanine dipeptide is taken as
$\mathcal P(\varphi,\psi) = (\cos(\varphi), \sin(\varphi), \cos(\psi), \sin(\psi))$,
so the periodic condition of the FES is guaranteed.
Similarly, the preprocessing operator for the alanine tripeptide is
$\tilde{\mathcal P}(\varphi_0, \psi_0, \varphi_1, \psi_1) =
(\mathcal P(\varphi_0, \psi_0), \mathcal P( \varphi_1, \psi_1))$.
Model training is carried out under the deep learning framework TensorFlow~\cite{abadi2016tensorflow}, 
using the Adam stochastic gradient descent algorithm~\cite{kingma2014adam} with a batch size of $\vert B\vert = 20$. 
The learning rate is 0.001 in the beginning and decays exponentially according to
$  r_l(t) = r_l(0) \times d_r ^{t / d_s}$,
where $t$ is the training step, $d_r = 0.96$  is the decay rate,  
and $d_s = 50 \times \vert D\vert / \vert B\vert$ is the decay step.
\recheck{
The total number of training steps is $12500\times\vert D\vert/\vert B\vert$. Currently,  the DNN structure and 
hyperparameters in the training algorithm are decided empirically. 
Before performing the full reinforced dynamics, we typically accumulate some
 data from some small scale simulations, test the performance of different DNN models 
and training schemes, and then fix the optimal strategy in terms of 
accuracy and efficiency.
In practice, we find that a DNN model with a decreasing number of nodes going from 
the innermost to the outermost hidden layers performs better in our test cases.
}

In each reinforced dynamics step, four DNN models with independent random initialization are trained in the same way to 
compute the uncertainty indicator. 
The biased MD simulations of alanine dipeptide and tripeptide last for 100~ps and 140~ps, respectively.
The CV values along the MD trajectories are computed and recorded in every 0.2~ps.
We assume no \emph{a priori} information regarding the FES,
so a brute-force simulation is performed for the 0th iteration step (we count the iterations from 0).
In each iteration at most 50 recorded CV values in the region with high uncertainty are added to the training dataset $D$.
Restrained MD simulations with spring constant  $500\,\mathrm{kJ/mol/rad}^2$ are performed to estimate the mean forces by~Eq.~\eqref{eqn:mf-res}.
Each restrained MD simulation is $100$~ps and 140~ps long for the alanine dipeptide and tripeptide, respectively.
The CV values are recorded in every 0.01~ps along the restrained MD trajectory to estimate the mean forces.
Both of the alanine dipeptide and tripeptide examples are carried out
on a desktop computer with an Intel i7-3770 CPU and 32~GB memory.

\subsection{Free energy surface construction}

\begin{figure}
  \includegraphics[width=0.23\textwidth]{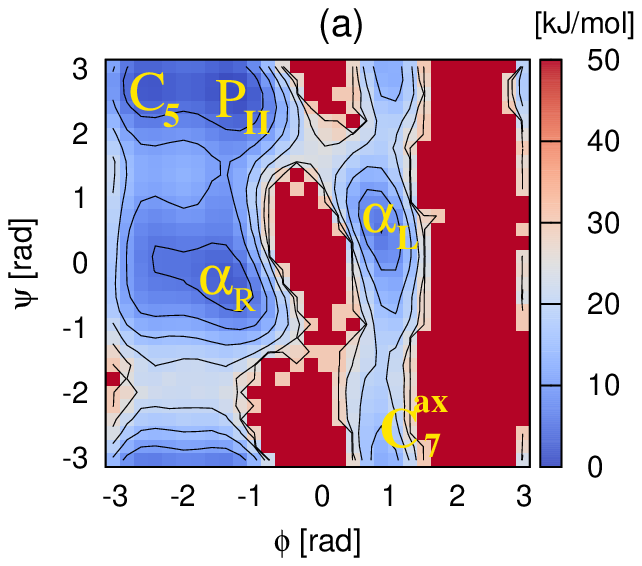}
  \includegraphics[width=0.23\textwidth]{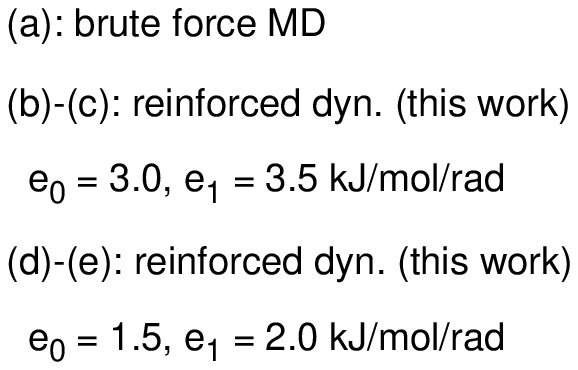}\\
  \includegraphics[width=0.23\textwidth]{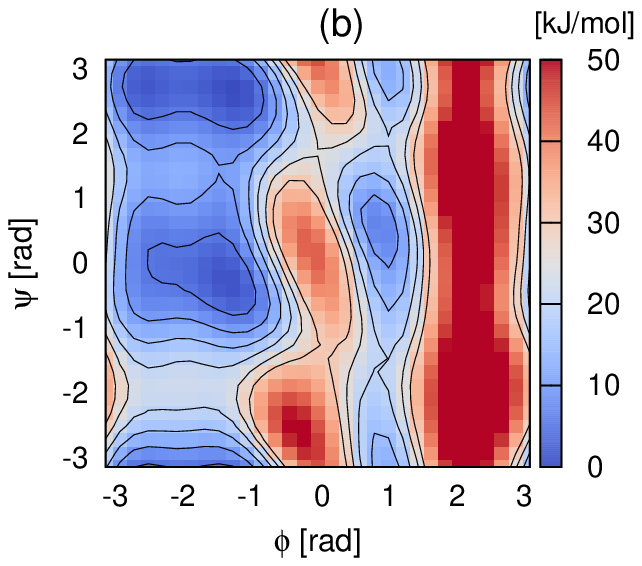} 
  \includegraphics[width=0.23\textwidth]{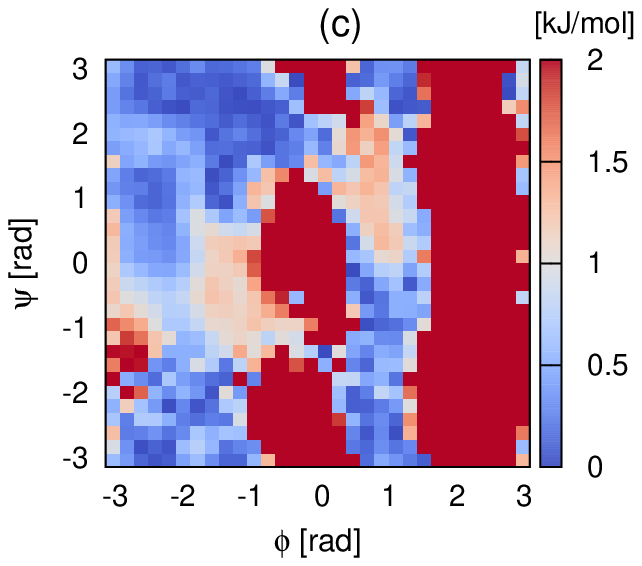} \\
  \includegraphics[width=0.23\textwidth]{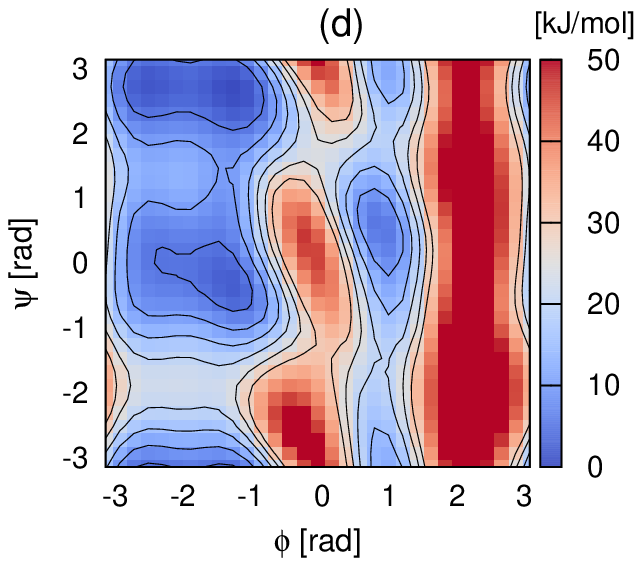} 
  \includegraphics[width=0.23\textwidth]{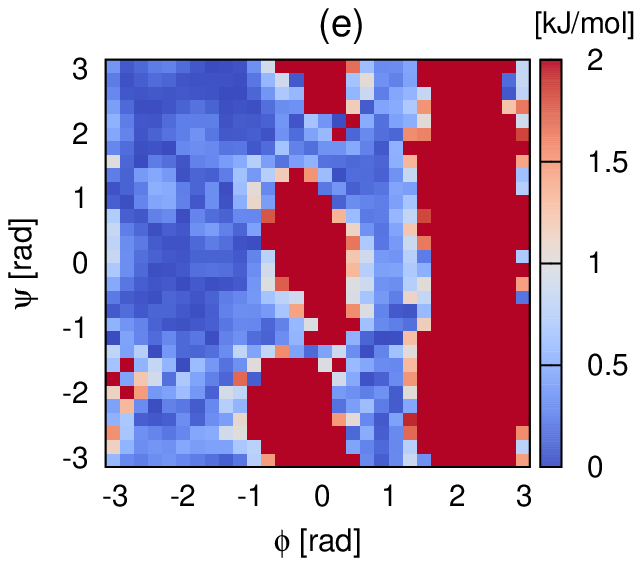} \\
  \caption{The free energy of alanine dipeptide on the $\varphi$-$\psi$ plane.
    The plots are obtained by
    (a) making log-scaled histogram of the CV values from brute-force MD simulations; 
    (b)--(c) using the reinforced dynamics  with uncertainty levels $e_0 = 3.0$, $e_1 = 3.5$~kJ/mol/rad. (b) plots the FES,  and (c) plots the error compared with the brute-force MD.
    (d)--(e) using the reinforced dynamics with uncertainty levels $e_0 = 1.5$, $e_1 = 2.0$~kJ/mol/rad. (d) plots the FES, and (e) plots the error compared with the brute-force MD.
    The contour lines in (a), (b) and (d) are plotted from 0~kJ/mol to 30~kJ/mol with an interval of 5~kJ/mol.
    The red regions in (a), (c) and (d) are the CV values that are never been visited by the MD trajectories.
    }
  \label{fig:ala-2-pp}
\end{figure}

The FES of the alanine dipeptide on the  $\varphi$-$\psi$ plane
(known as the Ramachandran plot)
is reported in Fig.~\ref{fig:ala-2-pp}.
We perform 6 independent brute-force MD simulations, with each $\sim$ 860~ns long,
thus in total 5.1~$\mu$s MD trajectories are used to estimate the FES and compare with the reinforced dynamics result.
The system has 5 metastable states \chem{\alpha_R}, \chem{C_5}, \chem{P_{II}}, \chem{\alpha_L} and \chem{C_7^{ax}}, as noted in Fig.~\ref{fig:ala-2-pp}~(a).
The \chem{C_5}, \chem{P_{II}} regions correspond to the dihedral angles observed in
the $\beta$-strands conformations.
The \chem{\alpha_R} and \chem{\alpha_L} regions correspond
to the dihedral angles of right- and left-handed \chem{\alpha}-helix conformations, respectively.
The transition between the \chem{P_{II}} and \chem{\alpha_L} has to
go over an energy barrier of $\sim$25~kJ/mol, or equivalently $\sim$$10 k_BT$.
The mean first passage time from the state \chem{P_{II}} to \chem{\alpha_L} is shown
to be 43~ns {for the same model~\cite{trendelkamp2016efficient}.}

In Fig.~\ref{fig:ala-2-pp}, the FES of alanine dipeptide sampled by the brute-force MD (a) is
compared with the one  constructed by reinforced dynamics  (b)
with uncertainty levels $e_0 = 3.0$~kJ/mol/rad and $e_1= 3.5$~kJ/mol/rad.
At the 9th iteration for (b), the biased simulation does not produce any CV value that belongs to the region with high uncertainty,
thus the computation stops.
In total (from the 0th to the 8th iteration)
198 CV values are added to the training dataset $D$ to train the FES.
It is observed that the reinforced dynamics is able to reproduce, with satisfactory accuracy, 
the FES at the important metastable states and transition paths of the system. 
The difference between (a) and (b) is plotted in (c).
The error of FES at states \chem{C_5}, \chem{P_{II}} and \chem{C_7^{ax}} is
below 0.5~kJ/mol,
while the error at \chem{\alpha_L} and \chem{\alpha_R} is around 1.5~kJ/mol.
The total biased MD simulation time is $10\times 0.1\,\mathrm{ns} = 1.0\,\mathrm{ns}$.
The total restrained MD simulation time is $198\times 0.1\,\mathrm{ns} = 19.8\,\mathrm{ns}$.
Thus the total MD simulation time is 20.8~ns,
which is only $\sim0.1$\% of the brute-force simulation length and half of the mean first passage time from \chem{P_{II}} to \chem{\alpha_L}
of the brute-force MD simulation.
\recheck{
The total wall time of all the trainings is $2.6\times 10^3$s,
while the total wall time of all the restrained MD simulations is $8.9\times 10^3$s.}

It is noted that the accuracy of the FES can be systematically improved
by using more strict uncertainty levels. 
The result of using $e_0 = 1.5$~kJ/mol/rad and $e_1= 2.0$~kJ/mol/rad is reported in Fig.~\ref{fig:ala-2-pp} (d) and (e).
In this case, the biased MD simulation does not generate CV values belonging to the region with 
high uncertainty at the 21st iteration.
In total (from the 0th to the 20th iteration) 303 CV values are added to the
training dataset $D$ to construct the FES.
The error of FES at all metastable states and transition regions
is uniformly below 0.5~kJ/mol. 
The total biased MD simulation time is $22\times 0.1\,\mathrm{ns} = 2.2\,\mathrm{ns}$.
The total restrained MD simulation time is $303\times 0.1\,\mathrm{ns} = 30.3\,\mathrm{ns}$.
Thus the total MD simulation time is 32.5~ns,
which is
50~\% longer than the reinforced dynamics with higher uncertainty levels ($e_0 = 3.0$~kJ/mol/rad and $e_1= 3.5$~kJ/mol/rad), but
still shorter than the mean first passage time from \chem{P_{II}} to \chem{\alpha_L}
of the brute-force simulation (43~ns).

\begin{figure}
  \centering
  \includegraphics[width=0.23\textwidth]{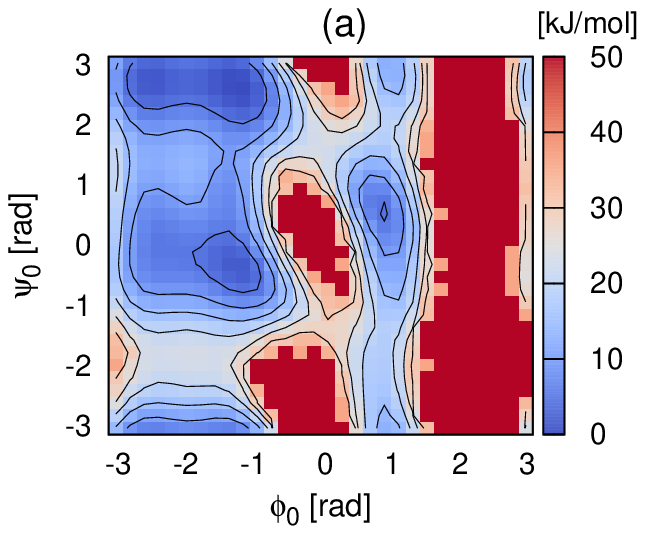}
  \includegraphics[width=0.23\textwidth]{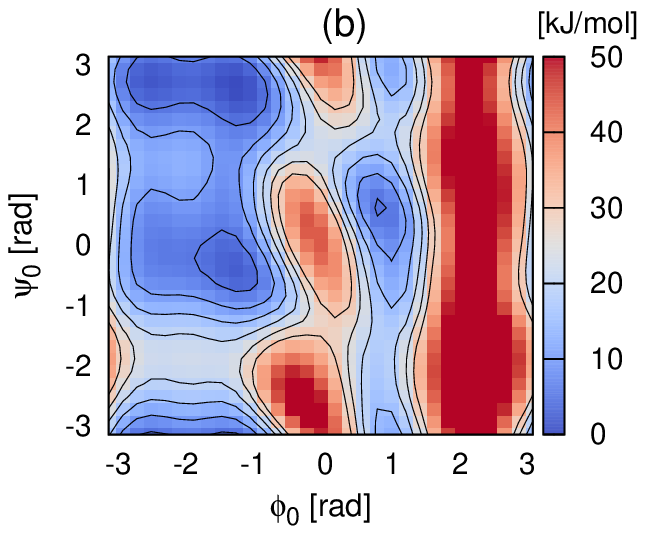}\\
  \includegraphics[width=0.23\textwidth]{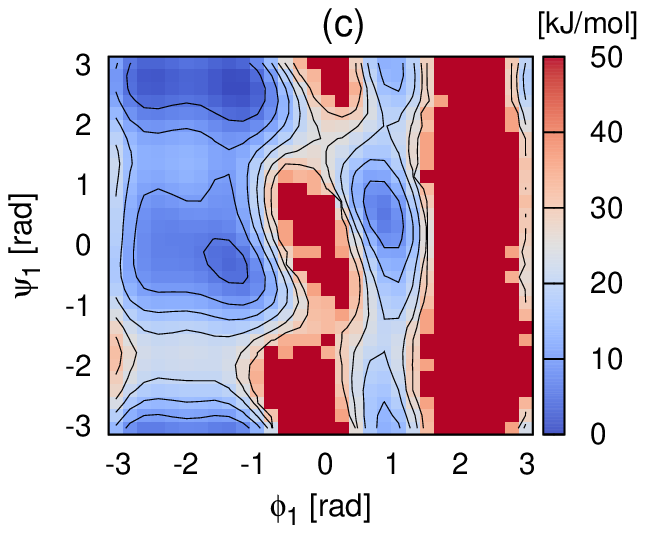}
  \includegraphics[width=0.23\textwidth]{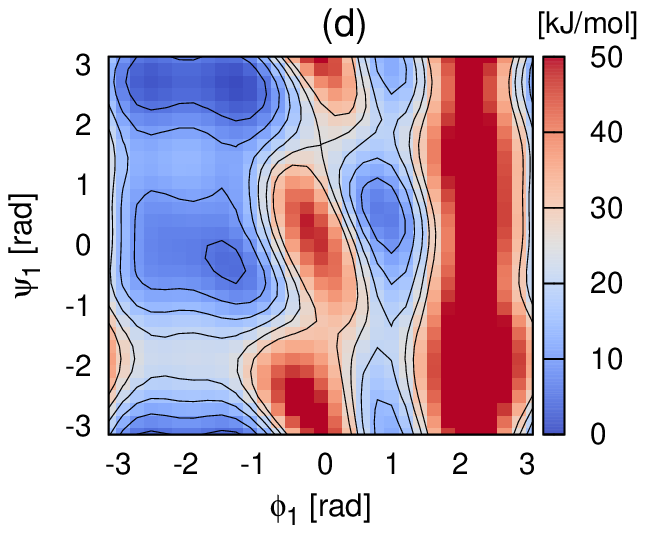}\\
  \includegraphics[width=0.23\textwidth]{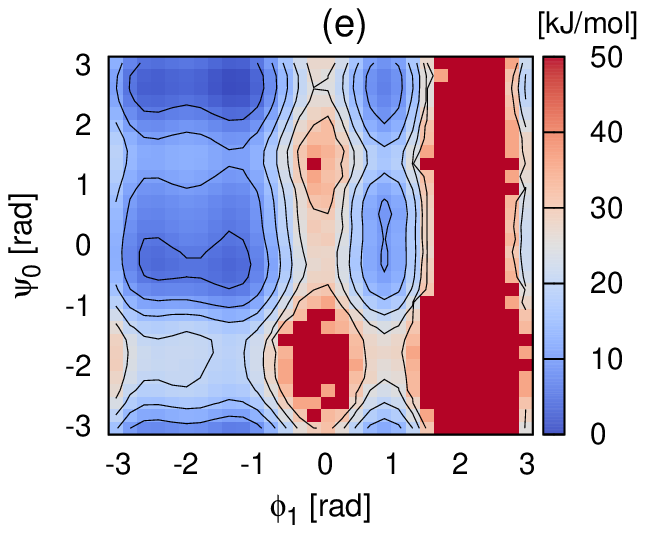}
  \includegraphics[width=0.23\textwidth]{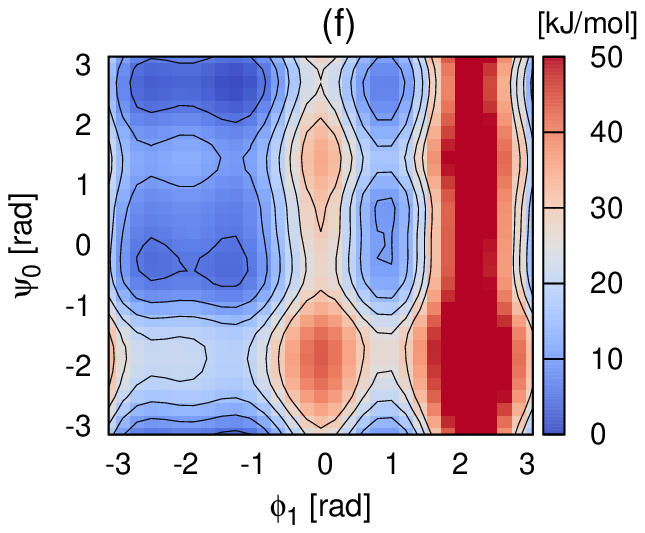}\\
  \caption{The free energy of alanine tripeptide projected on the $\varphi_0$-$\psi_0$ ((a) -- (b)),  $\varphi_1$-$\psi_1$ ((c) -- (d)) and $\varphi_1$-$\psi_0$ ((e) -- (f)) planes.
    The left column, (a), (c) and (e), are histogram plots of the CV values from brute-force MD simulations;
    The right column, (b), (d) and (f), presents the results of reinforced dynamics up to the 71st step, which is trained by 1363 CV values in the dataset $D$.
    }
  \label{fig:ala-3-proj}
\end{figure}

The information of the four-dimensional FES of the alanine tripeptide constructed by brute-force MD sampling and the reinforced dynamics
is presented in Fig.~\ref{fig:ala-3-proj}, by projecting on the $(\varphi_0, \psi_0)$, $(\varphi_1, \psi_1)$ and $(\varphi_1, \psi_0)$ planes.
For example,  the projection onto the $(\varphi_0, \varphi_0)$ variables is defined by
\begin{align}
  A(\varphi_0,\psi_0) =
  -\frac 1\beta\ln \iint d\varphi_1 d\psi_1
  e^{-\beta A(\varphi_0,\psi_0,\varphi_1,\psi_1)} + C,
\end{align}
where $C$ is a constant that is chosen to normalize the minimum value of $A(\varphi_0, \psi_0)$ to zero.
Projected free energies $A(\varphi_1,\psi_1)$ and $A(\varphi_1,\psi_0)$ are defined analogously.
The uncertainty levels of the reinforced dynamics are set to $e_0 = 3.0$~kJ/mol/rad and $e_1 =3.5$~kJ/mol/rad.
The biased MD simulation of the 72nd iteration does not find any CV value belonging to the region with
high uncertainty, so the process stops.
From the 0th to the 71st iteration, 1363 CV values are added to the training dataset $D$.
The total biased MD simulation time is $73 \times 0.14 = 10.22$~ns, 
while the total restrained MD simulation time is $1363 \times 0.14 = 190.82$~ns.
\recheck{
The total wall time of the restrained MD simulations is $6.2\times 10^4$s,
while the total wall time for training the networks is $1.1\times 10^5$s.}
For comparison, we carried out 18 independent brute-force MD simulation, each of which is 2.65~$\mu$s long,
so the total length of brute-force MD trajectories is 47.7~$\mu$s. 
Fig.~\ref{fig:ala-3-proj} shows that the reinforced dynamics  is able to reproduce the FES 
with satisfactory accuracy on all the projected planes.
It is noted that the projected FESs on both the 
$(\varphi_0, \psi_0)$ and $(\varphi_1, \psi_1)$ variables
are different from the FES of alanine dipeptide,
which indicates the correlation of backbone atoms.

\subsection{Illustration of the adaptive feature}

\begin{figure}[]
  \centering
  \includegraphics[width=0.49\textwidth]{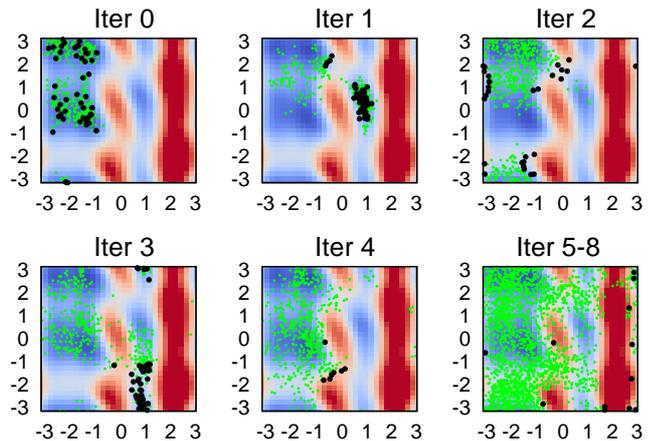}
  \caption{The CV values visited by the biased MD simulations (thin green dots) and those added to the training dataset $D$ (thick black dots)
    in each reinforced dynamics step.
    The uncertainty levels of the reinforced dynamics are set to $e_0 = 3.0$ and $e_1 = 3.5$~kJ/mol/rad.
    The color scale is the same as plot (b) of Fig.~\ref{fig:ala-2-pp}.
  }
  \label{fig:ala-2-iter}
\end{figure}

To highlight the adaptive feature of the reinforced dynamics,
we take the alanine dipeptide as an example, and illustrate in Fig.~\ref{fig:ala-2-iter}
the CV values visited in each biased MD simulation and those iteratively added to the training dataset $D$.
The uncertainty levels are $e_0 = 3.0$, $e_1 = 3.5$~kJ/mol/rad, and the reinforced dynamics stops at the 9th iteration.
In the 0th iteration, no \emph{a priori} information of the FES is available,
so the MD simulation is not biased.
The starting state of the simulation is \chem{P_{II}},
and the system spontaneously transforms to states \chem{C_5} and \chem{\alpha_R}
in the 0.1~ns simulation~\footnote{
The mean first passage time
from \chem{P_{II}} to \chem{C_5} and \chem{\alpha_R}
are 0.041 and 0.255~ns, respectively. See Ref.~\cite{trendelkamp2016efficient}.}, 
thus the visited CV values cover
\chem{P_{II}}, \chem{C_5} and \chem{\alpha_R}, 
and 50 of them are randomly chosen as training data.
The first DNN representation of FES is trained by these CV values, and is used to bias the system at the 1st iteration.
Since the first DNN representation is of good quality at states \chem{P_{II}}, \chem{C_5} and \chem{\alpha_R}, 
the system diffuses out of \chem{P_{II}}, \chem{C_5} and \chem{\alpha_R}, and is trapped by a new metastable state \chem{\alpha_L}.
Only the visited CV values that sample the metastable state \chem{\alpha_L} are added to the training dataset.
The DNN representation trained by the updated dataset is of good quality at states \chem{P_{II}}, \chem{C_5}, \chem{\alpha_R} and \chem{\alpha_L}.

Following this observation,
in the 2nd iteration, although the visited CV values cover a wide region including the metastable states \chem{P_{II}}, \chem{C_5}, \chem{\alpha_R} and \chem{\alpha_L},
only those in  the transition regions
between \chem{P_{II}} and \chem{\alpha_L}, and 
between \chem{\alpha_R} and \chem{P_{II}}/\chem{C_5} are added to the training set.
The CV values added in the 3rd iteration are those that 
sample the metastable state \chem{C_7^{ax}}
and the transition region between \chem{C_7^{ax}} and \chem{\alpha_L}.
The CV values added in the 4th iteration are those that sample
the transition region between \chem{C_7^{ax}} and \chem{\alpha_R}.

From the 5th to the 8th iteration, the DNN representation of the FES is of 
relatively good quality.
The CV values added to the training dataset are those that
sample the border of high energy peaks
at $\varphi \approx2$~rad and $\varphi \approx -0.5$~rad.
{At the 9th iteration, no CV value belonging to regions with high uncertainty is found
  because the pushing-back events happen so quickly
  that the CV values are not recorded by
the biased MD trajectory with the 0.2~ps recording interval.
  However, if we reduce the CV recording interval from 0.2~ps to 0.04~ps,
  19 CV values can still be identified to be in the regions with high uncertainty and used to start the next biasing-and-training iteration.
Since the construction of high energy FES peaks is of less interest, 
for the sake of computational cost, we do not use the smaller recording interval in our
result.
This means that the we ignore the
FES regions with sharp gradient so that
the biased system can only stay for a time scale
that is much shorter than the recording interval.
Better stopping criteria that guarantee the representation quality of 
the important structures of FES 
and excludes the irrelevant energy peaks are left for future studies.}

\subsection{Remark on the choice of CVs}
\begin{figure}
  \centering
  \includegraphics[width=0.155\textwidth]{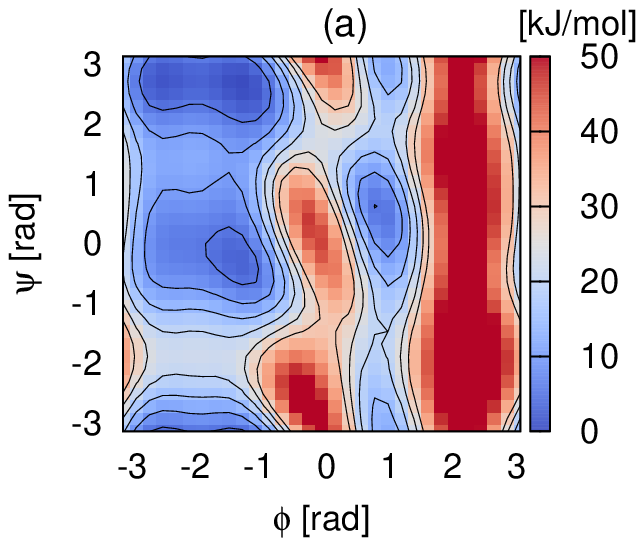}
  \includegraphics[width=0.155\textwidth]{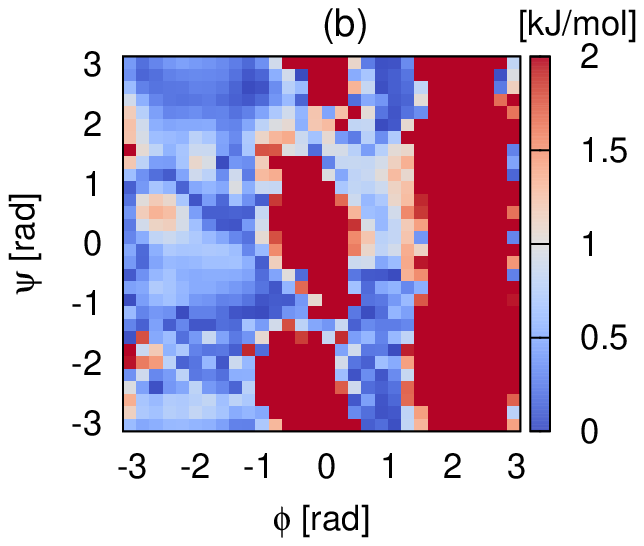}
  \includegraphics[width=0.155\textwidth]{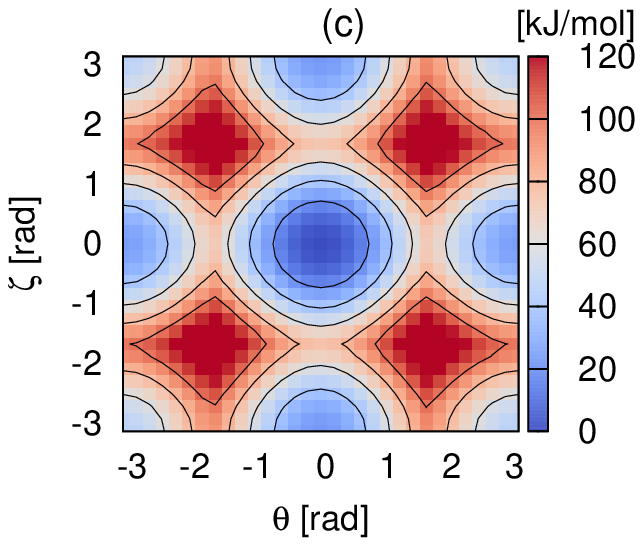}
  \caption{The FES of the alanine dipeptide computed in the CV space $(\varphi, \psi, \theta, \zeta)$.
    (a) projected on the $\phi$--$\psi$ plane, contour lines are plotted from 0~kJ/mol to 30~kJ/mol at an interval of 5~kJ/mol.
    (b) error of the $\phi$--$\psi$ projection compared to the brute-force simulation.
    (c) projected on the $\theta$--$\zeta$ plane, contour lines are plane from 0~kJ/mol to 100~kJ/mol at an interval of 20~kJ/mol.
    The uncertainty levels of the reinforced dynamics are set to $e_0 = 3.0$ and $e_1 = 3.5$~kJ/mol/rad.
  }
  \label{fig:ala-2-robust}
\end{figure}


One important issue is to find the right set of CVs  in order to capture the structural and dynamics
information that we are interested in.
However, this is a difficult problem and is not the topic of this work.
Here, we will study how the enhanced sampling and free-energy estimation are affected when 
(unnecessary) additional CVs are included.
We will see that the estimated free energy for the larger set of CVs is 
consistent with the one for the smaller set of CVs in the sense that after
projecting the former onto the smaller set of CVs, one recovers the latter.

To this end, we compute the FES of alanine dipeptide in a four-dimensional CV space
$(\varphi, \psi, \theta, \zeta)$ with two additional torsion angles
$\theta$ (O, C, N, \chem{C_{\alpha}}) and $\zeta$ (\chem{C_{\alpha}}, C, N, H).
The uncertainty levels that we use are $e_0 = 3.0$ and $e_1 = 3.5$~kJ/mol/rad.
The result is shown in Fig.~\ref{fig:ala-2-robust}.
The 4-dimensional FES projected on the $\theta$--$\zeta$ plane is shown in plot (c) of the figure.
The native state locates at $\theta = \zeta = 0$, while three metastable states are discovered at
$(\theta = 0, \vert\zeta\vert = \pi)$, $(\vert\theta\vert = \pi, \zeta = 0)$ and
$(\vert\theta\vert = \pi, \vert\zeta\vert = \pi)$.
They are denoted by $S_{00}$, $S_{01}$, $S_{10}$ and $S_{11}$, respectively.
The barrier between the native state $S_{00}$ and the metastable state $S_{01}$/$S_{10}$
is around 70~kJ/mol.
The free energy of metastable states $S_{01}$, $S_{10}$, and $S_{11}$ are 21~kJ/mol, 24~kJ/mol, and 46~kJ/mol, respectively, 
thus their contribution to the free energy projection on the $\phi$--$\psi$ plane is negligible.
A direct comparison of the free energy projection on the $\phi$--$\psi$ plane with the brute-force MD result
is shown in plot (b) of Fig.~\ref{fig:ala-2-robust}.
The error is less than 2~kJ/mol. 
The result is consistent with the $\phi$--$\psi$ free energy computed using reinforced dynamics (shown in Fig.~\ref{fig:ala-2-pp}).

\section{Application to polyalanine-10}
\label{sec:poly-ala-10}
In reinforced dynamics, both the neural network representation of the FES and the restrained simulation for mean forces are relatively
insensitive to the dimensionality of the CV space.
Thus it has the potential to be able to handle systems with a large set of CVs.
As an illustrative example, we investigate the metastable conformations of a 
polyalanine-10 (ACE-(ALA)\chem{_{10}}-NME) molecule.
In this example, rather than constructing an accurate free energy in the whole space of CVs, our goal is
 to efficiently search for the most stable structures in the conformational space of the system. 
We will demonstrate that reinforced dynamics allows us to explore very efficiently the 
most relevant metastable conformations of
this  molecule, including the $\alpha$-helix and $\beta$-strand conformations,
and to provide estimates for the relative stability between different metastable states.

One technical remark is that for computational efficiency, we adopt a multi-walker scheme of reinforced dynamics for this relatively high-dimensional case. 
In each iteration of this scheme, different walkers undergo 
biased dynamics independently under the same biased potential.
Next,  a set of CV values with high uncertainty are selected
and restrained simulations are performed to calculate the associated mean forces. 
Finally, the selected CV values and associated mean forces 
provided by all the walkers are merged and added to the dataset.
An ensemble of new neural network models are then trained with this larger dataset. 
The multi-walker scheme improves the efficiency of the data collection step and 
it helps to accelerate the exploration procedure.

\subsection{Simulation setup}
The system of polyalanine-10 (ACE-(ALA)\chem{_{10}}-NME) is
modeled by the Amber96 forcefield~\cite{kollman1996advances}.
The molecule is in the gas phase and is set in a $3.5~\textrm{nm}\times3.5~\textrm{nm}\times3.5~\textrm{nm}$ simulation region.
To start with, we prepare misfolded initial configurations of the molecule in three stages.
In the first stage, starting from an alpha-helix configuration,
two ends of the molecule is pulled along the $z$ direction in an extended simulation region ($3.5~\textrm{nm}\times3.5~\textrm{nm}\times30~\textrm{nm}$)
at rate 0.1~nm/ps for 100~ps.
During this process, no thermostat is used for the system.
At the end of this stage, the backbone of the molecule is fully extended, and the temperature of the system increases to 1155~K.
In the second stage, the pulling force is removed and the molecule is equilibrated at 300~K for 200~ps, using the velocity-rescaling
 thermostat~\cite{bussi2007canonical} with 0.2~ps of relaxation time
 and an integration time step of 1~fs.
In the third stage, an unbiased productive simulation of 200~ps is carried out at 
300~K with a time step of 2~fs.
100 candidate configurations along the trajectory of this simulation are saved in every other 2~ps.
Finally, 14 independent walkers are initialized with randomly chosen configurations from these candidates.

The torsion angles $\phi$ and $\psi$ associated to all the \chem{C_{\alpha}}s are used as CVs for the system,
thus the dimension of the CV space is 20.
The DNN model used in this example consists of 5 hidden layers of size $(M_1, M_2, M_3, M_4, M_5) = (360, 180, 90, 45, 20)$.
We found that the following procedure to be more efficient for the 
network training.
In the first 6 iterations, the weights in different DNN models are  randomly initialized, and are trained 
using the Adam stochastic gradient descent algorithm~\cite{kingma2014adam} with a batch size of $\vert B\vert = 64$. 
The learning rate $r_l$ is 0.003 in the beginning and decays exponentially according to
$  r_l(t) = r_l(0) \times d_r ^{t / d_s}$,
where $t$ is the training step, $d_r = 0.96$  is the decay rate,  
and $d_s = 10 \times \vert D\vert / \vert B\vert$ is the decay step.
The total number of training steps is $3000 \times \vert D\vert / \vert B\vert$.
After iteration 6, instead of randomly initializing the weights, we restart the training process using weights inherited from the previous iteration.
We use the same batch size and decay rate of the first 6 iterations, but use different learning rate of 0.0003 and decay step of $d_s = 5 \times \vert D\vert / \vert B\vert$.
This reduces the total number of training steps in each iteration to $1200 \times \vert D\vert / \vert B\vert$.
The biased MD simulations are 100~ps long.
The uncertainty levels are set to $e_0 = 6.0$ and $e_1 = 6.5$~kJ/mol/rad.
The CV values are computed and recorded in every 0.2~ps along the biased trajectories.
For each walker, at most 12 CV values in the region with high uncertainty are added to the training dataset $D$. 
For each added CV value, a $100$~ps restrained MD simulation is carried out, wherein the CV values are recorded in every 0.01~ps to estimate the mean forces using Eq.~\eqref{eqn:mf-res}.
The simulations are carried out on one cluster node with two Intel Xeon 
E5-2680 v4 CPUs and 64~GB memory.


\subsection{Structure optimization}
To find different metastable states and their relative stability, we combine the exploration stage, provided by the adaptively biasing procedure in reinforced dynamics,
with an optimization stage, which can be viewed as a postprocessing of the explored configurations.
In the exploration stage, due to the complexity of the 20-dimensional FES, we do not wait for the reinforced dynamics to stop by itself.
Instead, we stop the process at the 210th iteration.
The outputs of the biased MD simulations in each iterations, in total $14\times 211 = 2954$ configurations, are thus selected for the next stage.
We remark that basins associated to important metastable conformations may not be visited during the 210 iterations.
This seems to be a common issue of algorithms for conformation space exploration, no matter by enhanced sampling or by brute-force simulation.
However, reinforced dynamics drastically accelerates the efficiency of exploration and, due to the biasing procedure, new low-energy states are more likely to be explored in earlier iterations.
Although we stop the process at a certain number of iteration, further tests based on the accumulated dataset and restarted from the simulation can always be performed to check the results.
In the optimization stage, the 2954 configurations are first relaxed by brute-force MD for 200~ps.
Then the CV values corresponding to the relaxed configurations are taken as initial guesses
for the unconstrained minimization on the DNN represented FES, 
which is solved by the Broyden-Fletcher-Goldfarb-Shanno (known as BFGS) method~\cite{fletcher1987practical},
and the solutions are local minima of the FES.
The configurations are further relaxed with a restrained MD simulation centered at the corresponding local minima for 100~ps at a time step of 1~fs.

\begin{figure}
  \centering
  \includegraphics[width=0.48\textwidth]{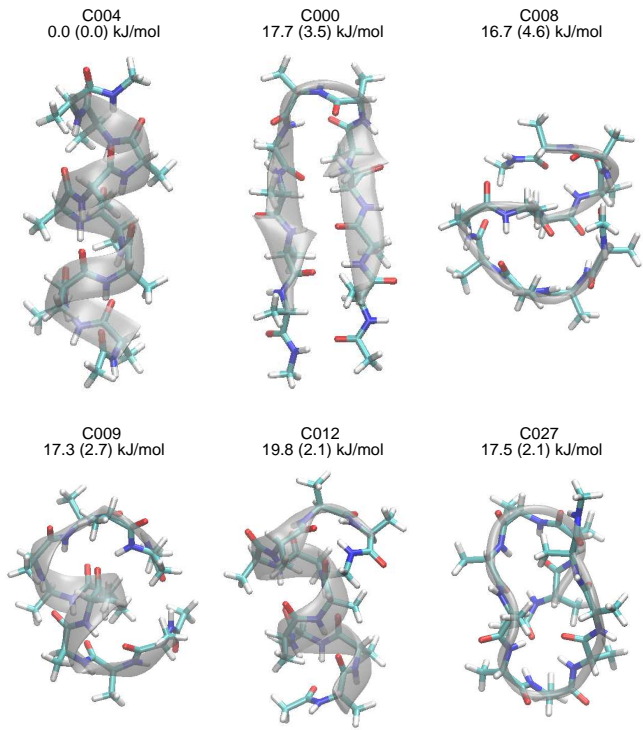}
  \caption{Schematic plot of the native state and five most stable conformations of the polyalanine-10 discovered by the reinforced dynamics.
    The gray shadows indicate the backbones of the conformations.
    Above each conformation, the cluster index (see text for details) and its free energy (in unit of kJ/mol) predicted by the reinforced dynamics are provided.
    The standard deviations of the free energy predictions are presented in the parentheses also in unit of kJ/mol.
  }
  \label{fig:ala-11-states}
\end{figure}

The local minimum with the lowest free energy corresponds to the native conformation,
which is the $\alpha$-helix conformation (see C004 in Fig.~\ref{fig:ala-11-states}).
The FES is thus shifted by the $\alpha$-helix free energy so that the global minimum takes the value of 0.
Among the 2954 configurations, 1047 configurations that have the free energy lower than 31.67~kJ/mol are collected. 
These configurations are clustered into 30 clusters
according to the root mean square deviation (RMSD) of \chem{C_{\alpha}}s by using the agglomerative clustering method with average-linkage criterion~\cite{scikit-learn}, 
and are coded as $\textrm{C000}, \textrm{C001}, \dots, \textrm{C029}$.
The largest averaged pairwise RMSD within one cluster is 0.86~\AA~(C003), which indicates a high conformational similarity within the clusters.
The configuration with the lowest free energy in one cluster is assigned to be the representative of that cluster, and its free energy is referred to as ``the free energy'' of the conformation.

The native conformation (C004) and
five metastable conformations with the lowest free energies are
presented in Fig.~\ref{fig:ala-11-states}.
Their relative stability with respect to the native state and the standard deviations of the free energy predictions are also presented in the figure.
The metastable conformation C000 corresponds to the $\beta$-strand conformation, while
the metastable conformations C008, C009, C012 and C027 are misfolded conformations.
The predicted free energies of the metastable conformations are very close,
thus considering the uncertainties in these free energies,
we can not tell whether one metastable state is more stable than another from the current reinforced dynamics simulation.


\begin{figure}[tb]
  \centering
  \includegraphics[width=0.4\textwidth]{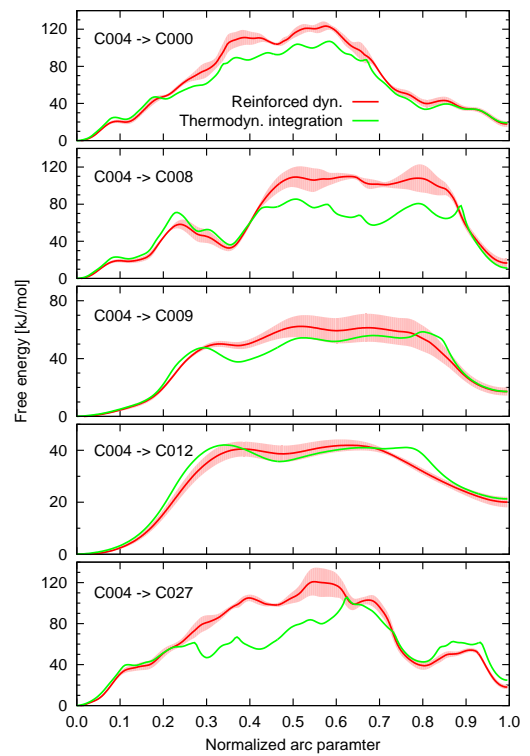}
  \caption{The free energy along the transition paths from the native to metastable conformations. 
    The transition paths are computed by the string method. 
    The free energies computed by the thermodynamic integration (green lines) and predicted by the reinforced dynamics (red lines) are demonstrated.
    The standard deviations of the free energy predictions are presented by red shadows.
  }
  \label{fig:ala-11-strm}
\end{figure}

We also computed the transition paths from the native state to the five metastable state using the string method~\cite{e2002string,e2007simplified}.
The strings are discretized by 224 nodes.
At each node a restrained MD of length 1600~ps is performed,
and the CV values are recorded every 0.01~ps to compute the mean force by Eq.~\eqref{eqn:mf-res}.
The free energies are then computed by using thermodynamic integration along the string (see the green lines in Fig.~\ref{fig:ala-11-strm}).
As a comparison,
the free energies predicted by the reinforced dynamics along the same paths  are plotted as the red lines,
 the standard deviations in the free energy model predictions are presented by the red shadows.
The  free energy predicted by the reinforced dynamics is in satisfactory agreement
 with the thermodynamic integration for the transitions
C004$\rightarrow$C000, C004$\rightarrow$C009 and C004$\rightarrow$C012.
The computation of the transition paths C004$\rightarrow$C009 and C004$\rightarrow$C012 are easier,
because the $\alpha$-helical segments in the conformations C009 and C012 make them closer to the native state.
It is also observed that the free energy barriers in transitions C004$\rightarrow$C009 and C004$\rightarrow$C012 are lower than others.
Along the paths C004$\rightarrow$C008 and C004$\rightarrow$C027,
the reinforced dynamics is quite accurate near the native and the metastable states.
However, in the middle section of the paths, there are  clear differences  from the result of the thermodynamic integration.
Many factors may contribute to this:
Between the native and a metastable state, there may exist multiple transition paths;
The path computed by the string method may not be the most probable path;
Some conformations along the path may not be well sampled by the reinforced dynamics.


\section{Conclusion and perspective}
In summary, reinforced dynamics is a promising tool for exploring the configuration space and calculating the free energy of atomistic systems. 
Even though we only presented examples of bio-molecules, it should be clear that the same strategy should also be applicable to many different tasks like studying the phase diagrams of condensed systems.
In particular, due to the ability of the deep neural networks in representing high dimensional functions \cite{han2017deep,zhang2017deep,schneider2017stochastic,lecun2015deep},
we expect the reinforced dynamics to be particularly powerful when the dimensionality of the CV space is high.
In addition, one should be able to  couple it with optimization algorithms in order to perform structural optimization.

\begin{acknowledgments}
We are grateful to Jiequn Han and Eric Vanden-Eijnden for their helpful comments.
We also thank Luca Maragliano for sharing with us the data of alanine dipeptide 
from the single-sweep method.
The work of L. Zhang and W. E is supported in part by ONR grant N00014-13-1-0338, DOE grants DE-SC0008626 and DE-SC0009248, 
and NSFC grants U1430237 and 91530322.
The work of H. Wang is supported by the National Science Foundation 
of China under Grants 11501039 and 91530322, 
the National Key Research and Development Program of China 
under Grants 2016YFB0201200 and 2016YFB0201203, 
and the Science Challenge Project No. JCKY2016212A502.
Part of the computational resources is provided by the Special Program for Applied Research 
on Super Computation of the NSFC-Guangdong Joint Fund under Grant No.U1501501.
\end{acknowledgments}


\end{document}